\documentstyle[12pt]{article}
\begin{document}

\begin{flushright}
NUB-TH-3151/97\\
CTP-TAMU-03/97\\
\end{flushright}
\begin{center}
{\bf Non-Universal Soft SUSY Breaking and Dark Matter}
\end{center}
 
\begin{center}
{Pran  Nath\\
Department of Physics, Northeastern University\\
Boston, MA  02115}\\
{R. Arnowitt\\
Center for Theoretical Physics, Department of Physics\\
Texas A \& M University, College Station, TX  77843-4242}\\
\end{center}

\begin{abstract}
 An analysis is given of the effects of 
non-universal soft SUSY breaking masses in the Higgs sector and in 
the third generation 
squark sector, and it is shown that they are highly 
coupled.  Analytic
expressions are obtained for their effects 
on the parameters $\mu,m_A$ and on the third generation squark masses.
Non-universality effects on dark matter event rates in neutralino-nucleus 
scattering are analysed. It is found that the effects 
 are maximal in the range  
 $m_{\tilde\chi_1}\leq 65$~GeV where the relic density
  is governed by the Z and Higgs poles. In this range
   the minimum event rates can be
  increased or decreased by factors of O(10) depending on the 
  sign of  non-universality. Above this range Landau 
  pole effects arising from the heavy top mass tend to 
  suppress the non-universality effects.
   The effect of  more precise measurements of cosmological 
  parameters on event rates, which is expected to occur in 
  the next round of COBE like sattelite
  experiments, is also investigated.
  Implications for the analysis for  dark matter searches  
  are discussed.
\end{abstract}  

\noindent
{\bf 1. Introduction}\\
Although there is currently  a great deal of evidence for the 
existence of dark matter in the universe\cite{smith}, 
the nature of such dark matter
is still uncertain. COBE  and other experiments have 
provided further insights into the nature of dark matter which strongly
suggests more than one component to its structure\cite{primack}. 
Thus the simplest 
types of models to fit the power spectrum seen by COBE and other experiments 
require two components, a hot component and a cold component. Several 
other possibilities have also been discussed such as those using a cosmological 
constant. 

 In our analysis here we shall adhere
 to the simple two component picture of dark matter, with a hot 
 component(which could be a light neutrino with mass in the range of a 
 few electron volts), and a cold component which we  assume is 
 the lightest supersymmetric neutral particle, the neutralino. (For a review
  of supersymmetric dark matter see ref.3). For the 
 purpose of our analysis we   assume a mix of hot
 and  cold dark matter in the ratio of 
$\Omega_{HDM}:\Omega_{CDM}=1:2$ (consistent with the COBE data), 
 where $\Omega_i=\rho_i/\rho_c$ where $\rho_i$ is the matter density 
 in the universe due to matter of type i and $\rho_c$ is the critical 
 matter density needed to close the universe. Assuming an inflationary 
 scenario with $\Omega =1$, the condition that $\Omega_B\leq 0.1$
 as implied by the Big Bang Nucleosynthesis, and h in the experimental range 
 $0.4\leq h \leq 0.8$ where h is the Hubble parameter in units of 100 km/s Mpc, 
 one finds $\Omega_{\tilde\chi_1} h^2$ in the range  

~\\
\begin{equation}
0.1\leq\Omega_{\tilde\chi_1}~h^2\leq 0.4
\end{equation}
~\\ 
$\Omega_{\tilde\chi_1} h^2$ is the quantity computed theoretically and thus the
above condition acts as the primary constraint on the neutralino dark matter
analysis. 	
We shall discuss in the latter part of this paper the effect of 
constraining the $\Omega_{\tilde\chi_1}h^2$ range more narrowly.

  The purpose of this paper is to analyse the effects of non-universal 
  boundary conditions
  on soft SUSY breaking masses at the GUT scale on event rates in 
  neutralino-nucleus scattering, as almost all of the previous analyses,
  with the exception of the work of Ref. (4), are in 
  the framework of supergravity unification  using only
  the universal boundary conditions. 
  
  The outline of  the paper is as follows: in Sec. 2 we give the salient
  points of supergravity unification which provides the framework of our
  analysis. In Sec. 3 we discuss non-universal soft breaking in the 
  Higgs sector and  in the third generation squark sector and show that their 
  effects
  at low energy are highly coupled. We then obtain solutions in a closed form at 
  the 
  electro-weak scale of the relevant low energy parameters that contain the
  non-universalities. In Sec. 4 we
  discuss the constraints that are imposed on the event rate
  analysis. In Sec. 5 we   
   give the basic formulae that enter in the analysis of scattering 
  of the neutralino by nuclear targets. In Sec. 6 results of the event
  rate analysis are given and a comparison made between the cases with
  universal and non-universal soft SUSY breaking boundary conditions 
  at the GUT scale. In Sec. 7 we discuss the effects of grand  unification
  on non-universal parameters.
   In Sec. 8 we discuss the effects that a more accurate
  determination of the Hubble parameter, which is expected to occur in the next 
  round of COBE like sattelite experiments, will have on the event rate
  analysis. Conclusions are given in Sec. 9. New results of this paper 
  are contained in Secs. 3,6,7,8, in Figs. 1-7, in Tables 1-5 and in 
  Appendix A.

 
\noindent
{\bf 2. Neutralino Dark Matter in Supergravity Grand Unification}\\
In the analysis of this paper we use the framework of supergravity
grand unification(SUGRA)\cite{cham1}. In this picture one assumes the existence of a 
supergravity grand unified theory which in its symmetric phase operates in the 
region between the string scale and the grand unification scale.
Supersymmetry is broken in the theory via a hidden sector. In the minimal
supergravity unification one finds that after breaking
of supersymmetry and of the grand unified symmetry (where one 
assumes that the grand unified theory breaks to the gauge group 
$ SU(3) \times SU(2)\times U(1)$ ) one has a low energy theory 
 (obtained by integrating out the heavy modes and the modes of the hidden
sector)  which can be given in terms of just five 
arbitrary parameters\cite{cham1}$^-$\cite{nac}. 

These consist of   the universal soft SUSY scalar mass at the GUT scale $m_0$,
the universal gaugino mass at the GUT scale $m_{1/2}$, the universal 
trilinear scalar coupling at the GUT scale $A_0$, $B_0$ which is the 
co-efficient of the bilinear term  $\mu_0H_1H_2$ at the GUT scale,  
and the Higgs mixing
parameter $\mu_0$ at the GUT scale. In addition there are the GUT parameters $M_G$ and $\alpha_G$
which are determined by the grand unification condition.

     A remarkable aspect of supergravity grand unification is the 
     breaking of the electro-weak symmetry by radiative effects.
     Radiative breaking is governed by the equations\cite{inoue}

\begin{equation}
\mu^{2}={{\mu_{1}^{2}-\mu_{2}^{2}tan^{2}\beta}\over{tan^{2}\beta -1}}-{1\over
2}M_{Z}^{2};~~ sin2\beta = {{-2B\mu}\over{2\mu^{2}+\mu_{1}^{2}+\mu_{2}^{2}}}\\
\end{equation}
     
\noindent
where B is the value of the parameter $B_0$ at the electro-weak scale,
  $\mu_{i}^{2}=m_{H_{i}}^{2}+\Sigma_{i}$ where $m_{H_{i}}^{2}$ is the
running $H_{i}$ mass at the scale Q $\approx M_{Z}$ and $\Sigma_{i}$ are loop
corrections\cite{ellis1,anloop}. $m_{H_{i}}^{2}$ are given by
\begin{equation}
m_{H_{1}}^{2}=m_{o}^{2}+m_{1/2}^{2} g(t)\\
\end{equation}
\begin{equation}
m_{H_{2}}^{2}=m_{1/2}^{2}e(t)+A_{o}m_{1/2}f(t)+m_{o}^{2}h(t)
-k(t)A_{o}^{2}\\
\end{equation}
\noindent                             
 Here t=ln($M_{G}^{2}/Q^{2}$), and   e,f,g,h,k are form factors 
 defined in Ref. (12).  On using the radiative breaking equations one can 
 determine the parameter $\mu$ from the Z boson mass and one can
 also eliminate the
 parameter B in favor of tan$\beta\equiv <H_2>/<H_1>$.
 Thus in the low energy domain, the theory can be described by 
 the parameters

\begin{equation}
m_0, m_{\tilde g}, A_t, tan\beta,sign(\mu)
\end{equation}
where $A_t$ is the value of $A_0$ at the electro-weak scale.
Further, at the electro-weak scale the $SU(3)\times SU(2)\times SU(1)$
 gaugino masses  are given by
$m_{i}=(\alpha_i/\alpha_{G})m_{1/2}$. ( For radiative corrections 
to this formula see Refs. (13-14)).
We note that the situation in minimal supergravity is in sharp contrast
to that for MSSM. In MSSM one has 110 parameters in the low energy domain.
 In contrast, in minimal supergravity unification
(MSGM) one has only four parameters (and one sign) as discussed above. The 
minimal model has 32 supersymmetric particles, and all their masses and 
coupling constants are determined by just the parameters of Eq. (5).
 
 One of the interesting features that emerges from supergravity 
 unification is that the model predicts as a consquence of its 
 dynamics which particle is the lowest lying supersymmetric particle(LSP). 
 One finds 
 that over almost all of the parameter space that the neutralino is the 
  LSP. Thus with the assumption of R parity invariance, one finds that
 supergravity unified theory provides a candidate for cold dark matter.
 The LSP neutralino is an admixture of four neutral states, i.e., of
 gauginos $\tilde{W}_{3},\tilde{B}$ and of 
  Higgsinos $\tilde{H}_{1}, \tilde{H}_{2}$. Denoting the lightest
  neutralino by  $\tilde{\chi}_{1}$ we may write 
\begin{equation}
\tilde{\chi}_{1}=n_{1}\tilde{W}_{3}+n_{2}\tilde{B}+n_{3}\hat{H}_{1}+n_{4}
\tilde{H}_{2}\\
\end{equation}
\noindent   
where the co-efficients $n_{i}$ are to be 
determined  by diagonalizing the neutralino mass matrix given in Appendix A. 
 An interesting property of  radiative breaking is that it produces 
 the phenomenon of scaling in a significant part of  the parameter 
 space\cite{scaling1}.
 In the region of scaling one has $\mu^{2}>>M_{Z}^{2}$.
 In this region the eigenvalues and the eigenvectors of the neutralino 
 mass matrix can be computed in a perturbative fashion by expanding in 
 powers of $M_Z/\mu$\cite{predictions}. 
The analysis shows that in the scaling limit the neutralino is mostly
a Bino with  $n_{2} > 0.95$. Further one finds that 
$n_{3}, n_{4}$ and $n_{1}$ are  typically of first order, i.e., $O(M_{Z}/\mu)$. 
 These expansions are useful  in understanding the relic density 
 and the event rates. We add here a note  of caution.
  Inspite of the fact that the Higgsino components 
 are generally small their effects on event rates are generally very
 significant. Thus large inaccuracies can result from a neglect of
 the Higgsino components.\\


\noindent
{\bf 3. Effects of Non-universal Soft SUSY Breaking}\\
In this section we want to study the  effects of non-universalities
in soft SUSY breaking 
on the low energy parameters. While most of the previous analyses
in supergravity grand unification have been carried out with universal
soft SUSY breaking, the general framework of the theory allows for
non-universal soft SUSY breaking terms. Thus, for example, under the assumption of a 
general Kahler potential, deviations from universalities can be 
generated\cite{soni}. Of course the non-universalities are severely
constrained,
most significantly from flavor changing neutral current(FCNC) constraints.
In the analysis here we shall limit the nature of non-universalities 
to the Higgs sector and to the third generation sector. Higgs sector 
non-universalities are not strongly constrained by 
FCNC and have been discussed in the literature\cite{matallio,olech,polon,berez}
 often in the context of certain
SO(10) models to achieve radiative breaking of the electro-weak symmetry.
In our analysis here we show that the non-universalities in the Higgs 
sector and the non-universalities in the third 
generation up squark sector are highly coupled  due to 
renormalization group effects below the GUT scale. 
It is convenient to parametrise the 
non-universalities in the Higgs sector by  $\delta_1,\delta_2$ 
and the non-universalities in the third generation squark sector by $\delta_3,
\delta_4$ so that at the GUT scale
 $M_G$ one has
~\\
\begin{equation}
m_{H_1}^2 =  m_0^2 (1 + \delta_1),  ~~m_{H_2}^2 = m_0^2 (1 + \delta_2)
\end{equation}
\begin{equation}
  m_{\tilde Q_L}^2=m_0^2(1+\delta_3), ~~m_{\tilde U_R}^2=m_0^2(1+\delta_4)
\end{equation}

\noindent
where $m_0$ is the universal scalar mass of the first two generation masses.
 We shall assume that the remaining
sectors of the theory are universal. 
We shall give an analytic solution to the non-universal 
effects on the mass spectra at low energy. 
 We begin by a discussion of 
 the  case of the ${H_1}$ mass and the masses of the first
two generations of sparticles. For $m_{H_1}^2$ we obtain
\begin{equation}
m_{H_{1}}^{2}=m_{o}^{2}(1+\delta_1)+m_{1/2}^{2} g(t)+\frac{3}{5}S_0p\\
\end{equation}
Here the last term arises from the $Tr(Ym^2)$ term in the renormalization 
group evolution equations\cite{mv}. This term vanishes in the universal case
because of the anomaly cancellation constraint $Tr(Y)=0$. However, for the
non-universal case it is non-vanishing and one has
\begin{equation}
S_0=Tr(Ym^2)=m_{H_2}^2-m_{H_1}^2+\sum_{i=1}^{n_g}(m_{\tilde q_L}^2-2 
m_{\tilde u_R}^2 +m_{\tilde d_R}^2 - m_{\tilde l_L}^2 + m_{\tilde e_R}^2) 
\end{equation}
where all the masses are at the GUT scale and
 $n_g$ is the number of generations. In Eq. (11) p is defined by 
\begin{equation}
p=\frac{5}{66}[1-(\frac{\tilde\alpha_1(t)}
{\tilde\alpha_1(0)})]
\end{equation}
where  $\tilde\alpha_1=g_1^2(0)/(4\pi)$ and  $g_1(0)$ is the
U(1) gauge coupling constant at the GUT scale. The corrections to the sparticle masses for the first 
two generations are listed in Appendix A. 
 We note that although the non-universalities at the GUT scale
in the first two generations vanish there are effects at the electro-weak
scale in the sparticle masses of the first two generations because of 
non-universalities in the Higgs sector and in the third generation 
via the trace anomaly term. To get an idea of the size of the effects 
from this term, for $M_G=10^{16.2}$ GeV and $\alpha_G=1/24$ one has
p=0.0446. Thus the effect of the trace term is order a few percent for
$|\delta_i|<1$. 

The analysis of non-universalities 
on the squark masses in the third  generation is more complicated. 
Here  $m_{H_{2}}^{2}$, $m_{\tilde U}^{2}$ and $m_{\tilde Q}^{2}$
obey the coupled renormalization group equations (61)-(63) of Appendix A.
 Solution to 
these gives for $m_{H_{2}}^{2}$ 
\begin{equation}
m_{H_{2}}^{2}=m_0^2\Delta_{H_2} +m_{1/2}^{2}e(t)+A_{o}m_{1/2}f(t)
+m_{o}^{2}h(t)-k(t)A_{o}^{2}-\frac{3}{5}S_0p
\end{equation}
where $\Delta_{H_2}$ is given by 
\begin{eqnarray}
\Delta_{H_2}=\frac{(D_0-1)}{2}(\delta_2+\delta_3+\delta_4)+ \delta_2
\end{eqnarray}
and $D_0$ defines the top Landau pole, 
i.e., 
\begin{equation}
y_0=\frac{y_t}{E(t)D_0}; D_0=1-6 y_t \frac{F(t)}{E(t)}
\end{equation}
Here $y_0=h_t^2/(4\pi)^2$ where $h_t$ is the top Yukawa coupling, and 
\begin{equation}
E(t)=(1+\beta_3(t))^{\frac{16}{3b_3}}(1+\beta_2(t))^{\frac{3}{b_2}}
(1+\beta_1(t))^{\frac{13}{9b_1}}
\end{equation} 
In Eq. (15)  $\beta_i=\alpha_i(0)b_i/4\pi$, where  $\alpha_i(0)=\alpha_G$ 
are the gauge
coupling constant co-efficients at the GUT scale 
($\alpha_1=(5/3)\alpha_Y$), $b_i$ are 
the one loop
beta function co-efficients defined  by $(b_1,b_2,b_3)=
(33/5,1,-3)$, 
and $F(t)=\int_0^t E(t)dt$. 
  The non-universality effects in the stop matrix enter via  $m_{\tilde Q}$
  and $m_{\tilde U}$ where  the stop mass matrix is 
\begin{equation}
\left(
{{ {m_{\tilde t_L}^2}\atop{-m_{t} (A_t + \mu ctn \beta)}}
{{-m_t (A_t +\mu ctn \beta)} \atop {m_{\tilde t_R}^2}   }}
\right)
\end{equation}
and $m^2_{\tilde t_{L}}$,  $m^2_{\tilde t_{R}}$ are given by 
\begin{equation}
m^2_{\tilde t_{L}}=m_{\tilde Q}^2+m_t^2+
(\frac{1}{2}-\frac{2}{3}sin^2\theta_W)M_Z^2cos2\beta 
\end{equation}
and 
\begin{equation}
m^2_{\tilde t_{R}}=m_{\tilde U}^2+m_t^2  
+(\frac{2}{3})sin^2\theta_w M_Z^2 cos2\beta
 \end{equation}
Here $m^2_{\tilde Q}$ and $m_{\tilde U}^2$ obey the set of coupled equations
given in Appendix A. With the assumption of non-universalities of Eqs. (7) and
(8) one has  for $m^2_{\tilde Q}$
\begin{eqnarray}
m^2_{\tilde Q}= m_0^2\Delta_{\tilde Q}+\frac{2}{3}m_0^2+\frac{1}{3}(m_0^2h-
kA_0^2+m_{\frac{1}{2}}
 A_0f+ m_{\frac{1}{2}}^2 e)\nonumber\\
 +\frac{\alpha_G}{4\pi}(\frac{8}{3}f_3+f_2-
 \frac{1}{15}f_1)m_{\frac{1}{2}}^2-\frac{1}{5}S_0p
 \end{eqnarray}
 where
\begin{equation}
\Delta_{\tilde Q} = \frac{(D_0-1)}{6}(\delta_2+\delta_3+\delta_4)
+\delta_3 
\end{equation}
Similarly for $m^2_{\tilde U}$ one has 
\begin{eqnarray}
m_{\tilde U}^2=m_0^2\Delta_{\tilde U}+\frac{1}{3}m_0^2+\frac{2}{3}
(m_0^2h-kA_0^2+m_{\frac{1}{2}}
 A_0f+ m_{\frac{1}{2}}^2 e)\nonumber\\
 +\frac{\alpha_G}{4\pi}(\frac{8}{3}f_3-f_2+\frac{1}{3}f_1)m_{1/2}^2
 +\frac{4}{5}S_0p 
 \end{eqnarray}
 where
\begin{equation}
\Delta _{\tilde U} = \frac{(D_0-1)}{3}(\delta_2+\delta_3+\delta_4)
+\delta_4
\end{equation}
In the above, the non-universalities are  all contained in the quantities 
$\Delta _{H_2}$, $\Delta _{\tilde Q}$, $\Delta _{\tilde U}$ and in the 
corrections involving the term $S_0$p. We note that because of a peculiar
accident that the sum of the corrections proportional to $S_0$p in the 
subsectors involving $H_2$,$\tilde Q$ and $\tilde U$ vanishes, the trace
anomaly receives no top quark Landau pole enhancement. A proof of this
result is given in Appendix A. An analysis similar to the above holds
in the bottom squark sector. Details are given
in Appendix A.\\

In addition to above, the quantities $\mu$ and $m_A$ are also effected. 
Using the radiative breaking relation Eq. (2) we find the following closed form
solution for  $\mu^2$ to one loop order:
\begin{equation}
\mu^2=m_0^2 C_1+A_0^2 C_2 +m_{\frac{1}{2}}^2C_3+m_{\frac{1}{2}}
A_0C_4-\frac{1}{2}M_Z^2+\frac{3}{5}\frac{t^2+1}{t^2-1}S_0p
\end{equation}
Here 
\begin{equation}
C_1=\frac{1}{t^2-1}(1-\frac{3 D_0-1}{2}t^2)+
\frac{1}{t^2-1}(\delta_1-\delta_2t^2-\frac{ D_0-1}{2}(\delta_2 +
\delta_3+\delta_4)t^2)
\end{equation}

\begin{equation}
C_2=-\frac{t^2}{t^2-1}k, 
~C_3=-\frac{1}{t^2-1}(g- t^2 e), 
~C_4=-\frac{t^2}{t^2-1}f 
\end{equation}
 where $t\equiv tan\beta$ and where the functions e,f,g,k are as defined 
in Eqs. (3) and (4). Similarly, for $m_A^2$ one has  
\begin{equation}
m_A^2=m_0^2 D_1+A_0^2 D_2 +m_{\frac{1}{2}}^2D_3+m_{\frac{1}{2}}
A_0D_4-\frac{1}{2}M_Z^2+\frac{6}{5}\frac{t^2+1}{t^2-1}S_0p
\end{equation}
where 
\begin{equation}
D_1=\frac{t^2+1}{t^2-1}(1-\frac{3 D_0-1}{2})+
\frac{t^2+1}{t^2-1}(\delta_1-\delta_2-\frac{ D_0-1}{2}
(\delta_2+\delta_3+\delta_4))
\end{equation}

\begin{equation}
D_2=-\frac{t^2+1}{t^2-1}k, 
~D_3=-\frac{t^2+1}{t^2-1}(g- e), 
~D_4=-\frac{t^2+1}{t^2-1}f 
\end{equation}
In the above we have assumed the existence of non-universalities on 
phenomenological grounds. Theoretically there are several possible 
sources from where such non-universalities can arise. One source  of
non-universalities is the general Kahler potential. In general the 
Kahler potential can have generational dependent couplings which lead
to  non-universalities in the visible sector after the breaking of supersymmetry 
in the hidden sector. However, even if the couplings in the Kahler 
potential were generation blind and the soft SUSY breaking masses
for the scalars were universal at a scale higher than the GUT scale  
one will have non-universalities
at the GUT scale  arising from renormalization group running of 
 the soft SUSY breaking parameters. 
Thus, for example, if one has universality of the soft parameters
at the string scale, the running of these parameters from the string 
scale down to the GUT scale will generate 
non-universalities at the GUT scale. Model calculations
indicate that such running typically generates 
$|\delta_{1,2}|\leq 0.5$, and a similar 
level of non-universality in the relevant mass spectra at 
the electro-weak scale is expected.  
However, enhancement of non-universalities can occur under special
circumstances.
 To show this enhancement we consider the expression for $\mu^2$. 
 Here the non-universalities  are all contained in $C_1$.
If the dominant universal terms cancel, then the non-universal effects
become large. This situation can  be easily seen to arise
 for large tan$\beta$.
Here  for  the case when $m_t\approx 167$ GeV,
the universal part cancels since 
$D_0\approx 1/3$   
and thus the non-universality effects get enhanced.
A similar enhancement of non-universalities also occurs from the $A_0$ 
dependence, although in a somewhat different manner. Here the enhancement 
occurs when the residue of the Landau pole in $A_0$ vanishes. All three
quantities, $m_{H_2}^2$, $m_{\tilde Q}^2$ and $m_{\tilde U}^2$, depend
on $A_0$ in the combination $kA_0^2-fA_0m_{1/2}$. Using the 
renormalization group equation for $A_0$ 
\begin{equation}
A_0=\frac{A_R}{D_0}-\frac{H_3}{F}m_{1/2};~~H_3=tE-F
\end{equation}
where $A_R=At-m_{1/2}(H_2-H_3/F)$ with $H_2$ defined in Ref. (12), one 
finds  that
\begin{equation}
 kA_0^2-fA_0m_{1/2}=\frac{1}{2}(1-D_0)\frac{A_R^2}{D_0}+\frac{1}{2}D_0(1-D_0)
 (\frac{H_3}{F})^2 m_{1/2}^2
\end{equation}
 The Landau pole contribution vanishes when $A_R=0$ which occurs when 
 $A_t\approx 0.61m_{\tilde g}$\cite{nwa}.
We give a graphical description of this phenomenon in Fig. 1 where 
$\mu$ is plotted as a function of $A_t$. For the specific set of 
parameters chosen in Fig. 1 the residue of the Landau pole vanishes
when $A_t/m_0\approx 0.7$. One finds that Landau pole effects are 
suppressed close to this value of $A_t$, and there is considerable
dispersion  among the three curves corresponding to the sets of
values $\delta_1=0=\delta_2$, $\delta_1=1=-\delta_2$ and $\delta_1
=-1=\delta_2$. However, away from the region of $A_t/m_0\approx 0.7$ 
one finds that the Landau pole effects begin
to dominate and diminish the effect of non-universalities.

\noindent
{\bf 4. Constraints on Dark Matter}\\
We explore the parameter space of supergravity unified models 
under the naturalness constraints which we assume to imply

\begin{eqnarray}
m_0\leq 1~ TeV, ~m_{\tilde g}\leq~ 1TeV,~tan\beta\leq 25
\end{eqnarray}
The constraints of electro-weak breaking and the condition that there 
be no tachyons then imply the following range for $A_t$\cite{nwa,predictions}:

\begin{equation}
 A_t/m_0 \geq -0.5
\end {equation}
 Constraints also arise from imposition of the condition that
there be no nearby minima with lower energies that break 
color and charge conservation\cite{ccb}(CCB constraints), and from 
experimental lower limits
on the SUSY mass spectra given by CDF, D0  and LEP experiments.

In addition to the above there is an important constraint that  arises
from the  decay $b\rightarrow s+\gamma$. 
 This decay proceeds via loop corrections and is thus sensitive 
to physics beyond the Standard Model(SM).
For the case of the SM, the loop contribution involves 
a W-t exchange, while for the supersymmetric case one has additional 
contributions arising from $W-H^-$,
$\tilde W -\tilde t$,  and  $\tilde Z -\tilde b$ exchanges. Thus
SUSY contributions are a priori as important as the SM contributions,
 and so one expects $b\rightarrow s+\gamma$ decay to act as an 
  important constraint on the parameter space of supergravity models. 
 The CLEO  value for  the branching ratio for 
$B\rightarrow X_s+\gamma$ is \cite{alam} 

\begin{equation}
BR(B\rightarrow X_s\gamma)=(2.32\pm 0.57\pm 0.35)\times 10^{-4}
\end{equation}

\noindent
If one  combines the errors in quadrature then  BR(B$\rightarrow
X_s\gamma)\cong(2.32\pm 0.67)\times 10^{-4}$. 
The SM prediction gives 
$BR[B\rightarrow X_s\gamma]\cong (3.28\pm
0.33)\times 10^{-4}$ for m$_t$ = 174 GeV\cite{bertolini,buras,misiak}. 
In supersymmety, theoretical analyses  
when subject to the current experimental limits,  impose serious constraints 
on the parameter space of the theory. 
 The parameter
space after the above constraints are imposed must still 
 be subject to  the constraint of 
relic density of Eq. (1). The quantity    $\Omega_{\tilde\chi_1^0}~h^2$
is computed theoretically in each point in the supergravity parameter 
space using the relation\cite{lee,kolb}
 ~\\
\begin{equation}
\Omega_{\tilde\chi_1^0} h^2\cong 2.48\times 10^{-11}{\biggl (
{{T_{\tilde\chi_1^0}}\over {T_{\gamma}}}\biggr )^3} {\biggl ( {T_{\gamma}\over
2.73} \biggr)^3} {N_f^{1/2}\over J ( x_f )}
\end{equation}
~\\
\noindent
where $ T_{\gamma}$ is the current background temperature, $T_f$ is the 
freezeout temperature,
 $(T_{\tilde\chi_1^0}/T_{\gamma})^3$ is the reheating 
factor, 
$N_f$ is number of
massless degrees of freedom at freezeout, $x_f= kT_f/m_{\tilde{\chi}_{1}}$,
  and 
~\\
\begin{equation}
J~ (x_f) = \int^{x_f}_0 dx ~ \langle~ \sigma \upsilon~ \rangle ~ (x) GeV^{-2}
\end{equation}
~\\
\noindent
Here $<\sigma v>$ stands for the thermal everage, where $\sigma$ is the
 the neutralino 
annihilation cross-section and v is the neutralino relative
velocity. In the analysis of the relic density we have used the
accurate method for its computation\cite{greist,accurate,accurate1,accurate2}
which carries out
a correct thermal averaging over the Z and the Higgs poles which appear
in the s channel in the neutralino annihilation. The accurate 
method for the computation of relic density is important 
as it has significant effects in the event rate analysis.
The relic density computed in the above fashion is then subject
to the constraints of Eq. (1) which further limits the parameter
space of the theory.\\

\noindent
{\bf 5. Detection of Neutralino Dark Matter}
We discuss now the  possibilities for the detection of the neutralino
dark matter.
Several techniques, both direct and indirect,
 have been discussed over the years for  its detection.The direct method
 includes (i)  scattering by nuclei\cite{goodman}-\cite{decarlos} in
 terestial detectors, 
 (ii) scattering by bound electrons\cite{starkman},
 (iii) use of overheated microbubbles\cite{collar}, 
 while the indirect methods include,
 (iv) annihilation in the center of sun and earth\cite{kamio1,kamio2}, 
 and (v) annihilation
 in the halo of galaxies. 
 For our analysis here we 
 limit ourselves to consideration of case (i).
 
 	 The total interaction in the scattering of neutralinos by 
 	 quarks is given by

\begin{equation}
L_{eff}= ({\bar{\tilde\chi}_1}{\gamma^{\mu}}\gamma_5\tilde\chi_1)[\bar{q}\gamma_{\mu}
(A_LP_L+A_RP_R)q]+(\bar{\tilde\chi}_1 \tilde\chi_1)(\bar{q}C m_qq) 
\end{equation} 
where $P_{R,L}=(1\pm\gamma_5)/2$. The part with terms $A_L,A_R$ is the
spin dependent(SD) part while the remainder is the spin 
independent(SI) part.
\noindent
The total event rate can thus be written as follows:
\begin{equation}
R=\left[ R_{SI}+R_{SD}\right ] \left [{\rho_{\tilde{\chi}_{1}}\over 0.3GeV
cm^{-3}}\right ] \left [{v_{\tilde{\chi}_{1}}\over 320 km/s}\right ]{events\over
kg~ da}  
\end{equation}
\noindent 
 where  $\rho_{\tilde\chi_1}$ is the local mass density of
${\tilde{\chi}_{1}}$ incident on the detector, and
$v_{\tilde{\chi}_{1}}$ is the incident ${\tilde{\chi}_{1}}$ 
velocity. $ R_{SI}$, is given by  

\begin{equation}
R_{SI}={16m_{\tilde{\chi}_{1}}M_N^3M_Z^4\over\left
[{M_N+m_{\tilde{\chi}_{1}}}\right ]^2}{\mid A_{SI}\mid^2}  
\end{equation}
 and the  spin dependent rate is given by

\begin{equation}
R_{SD} = {16 m_{\tilde{\chi}_{1}}M_N\over \left [M_N+m_{\tilde{\chi}_{1}}\right ]^2}
\lambda^2J(J+1)\mid A_{SD}\mid^2
\end{equation}
\noindent
where J is the nuclear spin and $\lambda$ is defined by $<N\mid\sum
{\stackrel{\rightarrow}{S}_i}\mid N>=\lambda<N\mid{\stackrel{\rightarrow}
{J}}\mid N>$. $A_{SI}$ and $A_{SD}$ are the corresponding amplitudes. 
We note that for large M$_N$,
 R$_{SI}\sim M_N$ while 
 $R_{SD}\sim 1/{M_N}$. These results imply that 
 the spin independent scattering becomes more 
 dominant as the nucleus becomes heavier.
 Both R$_{SI}$ and R$_{SD}$ contain theoretical and experimental 
 uncertainties. A major source of uncertainty in R$_{SI}$ 
 arises  from  the uncertainty in the matrix elements 
 of the operator $m_q\bar qq$ between
 nucleon states. There can be as much as 50\% uncertainty in the strange 
 quark contributions\cite{cheng} leading to a 30\% uncertainty in $R_{SI}$. 
 For the case of R$_{SD}$ the
 major source of ambiguity arises  from the matrix elements of the
 axial current between nucleon states, i.e., of $\Delta q$ 
 defined by $<n|\bar q\gamma^{\mu}\gamma_5q|n>$=2$s^{\mu}_n$$\Delta q$,
 where $s^{\mu}_n$ is the nucleon spin 4-vector. There exist two sets of
 determinations of $\Delta q$, an old determination using the 
 EMC data\cite{emc} and  a  new determination using  
 the SMC data\cite{ellis2}. For most nuclei the difference
 is not major and in  our analysis here
 we use the values of $\Delta q$ from the newer determination\cite{ellis2}. 
 In addition both rates have uncertainties due to the nature of the nuclear
 form factors. Thus theoretical predictions are accurate to within perhaps 
 a factor of two.\\

              
\noindent
{\bf 6. Event Rates with Universal and Non-universal Soft Breaking}\\
We give now an analysis of the event rates and draw a comparison between
the case where one uses universal boundary conditions at the GUT scale
for the soft SUSY breaking parameters and the case when on has non-universal
boundary conditions at the GUT scale. We begin with a discussion of event
rates when one includes non-universalities in the Higgs sector, i.e., 
when $\delta_3=0=\delta_4$. From Eqs. (23) and (24) one can see that 
a positive $\delta_1$ and a negative $\delta_2$ make a positive contribution
to $\mu^2$ while a negative $\delta_1$ and a positive $\delta_2$ make 
a negative contribution to $\mu^2$. If one limits $\delta_1$ and $\delta_2$
so that $|\delta_i|\leq 1$(i=1,2),  then the case $\delta_1=1=-\delta_2$
gives the largest positive contribution to $\mu^2$ while the case
$\delta_1=-1=-\delta_2$ gives the largest negative contribution to 
$\mu^2$. These cases represent the extreme limits of how large the 
non-universality efffects can be within our prescribed limits on
$\delta_i$.  With the above in mind we focus on three cases to get 
an idea of the  differences in the event rates between the universal
and the non-universal cases:\\
 	(i)$\delta_1=0=\delta_2$,\\
 	(ii)  $\delta_1=-1=-\delta_2$, and\\
	(iii)	$\delta_1=1=-\delta_2$\\
In Fig. 2 we exhibit the 
maximum and minimum of event rates for xenon 
 for the three cases listed above for $\mu>0$.The analysis is done
 	 by mapping the full parameter space 
	limited only by the naturalness constraints of Eq. (29), the 
	constraints  on the sparticle masses from CDF, DO and LEP 
	experiments, the 
	relic density constraint of Eq. (1) and  the experimental constraint
	from $b\rightarrow s+\gamma$ of Eq. (31). The dips in the 
	region below $m_{\tilde\chi_1}\leq 65$~GeV arise due to the rapid 
	annihilation in the vicinity of the Z pole and the Higgs pole. 
        We note  that the accurate
  	method mentioned earlier for the computation of the relic density
  	is important for the correct analysis of the  event rates in 
  	this region.
  	Comparison of cases(i) and (ii) 
	shows that the event rates of (i) below the region  
	$m_{\tilde\chi_1}\leq 65$~GeV can be enhanced by a factor of O(10)
	or more in the
	minimum curves for case(ii).This effect can be understood as
	follows: for case(ii) one has $\delta_1<0$ and $\delta_2>0$,
	which according to Eq. (23) drives $\mu^2$ towards the limit
	$\mu^2<0$ eliminating such points from the parameter space.
	Now this effect is more important for small tan$\beta$. Since
	small tan$\beta$ governs the minimum event rates one finds that
	in this case since small tan$\beta$ tend to get eliminated one  
	has the effect that the lower limit of the event rates 
	are raised.
	In contrast, for case(iii) one  has $\delta_1>0 $ and $\delta_2<0$
	which gives the opposite effect as expected, i.e., on finds 
	that the minimum event rates are reduced by a factor of O(10).
	However, above $m_{\tilde\chi_1}\geq 65$ the non-universalities do
	not dramatically affect the event rates and
	the event rates have 
	a uniform fall off as a function of the  neutralino mass
	for all the three cases. A similar effect occurs for
	  larger values of the top quark mass. However, here one finds that 
	  a larger value of the top mass gives a larger Landau pole 
	  contribution which diminishes the relative contribution of 
	  the non-universal terms.

	We note that the spin independent interaction contributes significantly
	more than the spin dependent intereaction. This result holds
	not only for a heavy target such as xenon but also for retatively
	light targets such as CaF$_2$. In Tables 1, 2 and 3 we give a
	comparison of event rates for six additional target materials
	, i.e., He, CaF$_2$, GaAs, Ge, NaI, and Pb, for the universal
	and non-universal soft SUSY breaking in the Higgs sector.
	(Recall, however, that there are significant uncertainties 
	discussed above in the predictions.) 
	Phenonmena similar to the ones discussed for xenon also appear
	for these cases.  
 	    Many of these materials listed in Tables 1, 2 and 3
 	    are  already being used as targets in the dark
	    matter detectors.  We note that for all the targets considered
	    the event rates 
	    span several decades in magnitude within the allowed 
	    domain of the parameter space. Thus a full exploration of the
	    parameter space certainly poses  a formidable challenge 
	    to experimentalists.\\
	    
	     The analysis we have given so far
	    is for the case $\mu>0$. As discussed earlier for the case $\mu<0$ 
	    a large part of the parameter space is eliminated because
	    of the  $b\rightarrow s+\gamma$ constraint. Consequently the 
	    maximum event rates for the $\mu<0$ cases are much smaller 
	    than for the $\mu>0$ case. A comparison of the allowed
	    range of event rates for xenon for the cases $\mu>0$ and
	    $\mu<0$ is given in Table 4 ($\mu>0$)  and Table 5 ($\mu<0$). 
	    One  finds that the event rates for the $\mu<0$ case are 
	    typically O($10^{-2}-10^{-3}$) smaller than for the $\mu>0$
	    case. Another important phenonmenon is that for
	    both signs of $\mu$ the parameter space corresponding to
	    $A_t/m_0<-0.5$ is eliminated.
	   The reason that positive values of $A_t$ are
	    eliminated arises from the fact that the lightest stop mass
	    turns tachyonic due to Landau pole effects, since the residue 
	    of the Landau pole $A_R^2=(A_t-0.61 m_{\tilde g})^2$ 
	    becomes large.\\
	    
	     In the analysis of this section thus far we assumed universality in
	     the third generation sector and examined only the effects of
	      non-universality in the Higgs sector. However, 
	       as discussed in Sec. 3 the non-universalities in the 
	     Higgs sector and in the third generation sector are  highly
	     coupled. Therefore,
	       we turn now to a 
	     discussion of what we might expect, from the non-universality
	     in the third generation sector. To get an idea of the nature
	     of corrections one might expect, we display below the 
	     numerical sizes of the non-universal corrections 
	     	     $\Delta_{H_2}, \Delta_{\tilde t_L},\Delta_{\tilde t_R}$.
	     For values $M_G=10^{16.2}$~GeV, $\alpha_G=1/24$ and
	     $m_t=175 ~GeV$ one has $D_0\simeq 0.27$ and from Eqs. (13), (20) 
	     and (22) one finds 	     
	     \begin{equation}
	      \Delta_{H_2}\simeq 0.64\delta_2-0.36(\delta_3+\delta_4)
	      \end{equation}
	     \begin{equation}
	       \Delta_{\tilde t_L}\simeq 0.88\delta_3-0.12(\delta_2+\delta_4)
	      \end{equation}
	     \begin{equation}
	      \Delta_{\tilde t_R}\simeq 0.76\delta_4-0.24(\delta_2+\delta_3)
	      \end{equation}
	     From Eq. (40) we find that 
	     a positive  $\delta_2$ in $\Delta_{H_2}$ can be simulated by a  
	     negative value of $\delta_3$ or a negative value of $\delta_4$,
	     and a converse situation occurs for a negative value of 
	     $\delta_2$. 
	     The effects of $\delta_3$ and
	     $\delta_4$ on  $\Delta_{\tilde t_L},\Delta_{\tilde t_R}$
	     are more complicated since  $\delta_3, \delta_4$ enter
	    $\Delta_{\tilde t_L},\Delta_{\tilde t_R}$ with opposite signs.     
	     The stop masses do not enter
	     directly in any significant manner in the amplitudes in the
	     neutralino  mass  range we investigate, although
	     they do affect the event rates by affecting the parameter space.
	     In spite of this complexity we can glean the pattern in which
	        $\delta_3,\delta_4$ affect the event rates by examining
	        the way they enter $\Delta_{H_2}$.  
	    In Fig. 3 we exhibit the event rates for xenon when 
	    $\delta_1=\delta_2=\delta_4=0$ and $\delta_3$ takes on a
	    range of values. The solid curve in Fig. 3 is for the case
	    $\delta_3=0$, the dotted curve for the case 
	    $\delta_3=1$, and the dashed for the case $\delta_3=-1$.
	     Fig. 3 shows that the minimum event rates 
	     in the region below 
	     $m_{\tilde\chi_1}<65$ GeV are lower relative to universal case
	     for the case  $\delta_3>0$. This is as
	     expected from our general analysis above, i.e., it is similar
	     to the case $\delta_2<0$ in Fig. 2.  Similarly the minimum
	     event rates in the region $m_{\tilde\chi}<65$ GeV are enhanced
	      relative to the universal case for $\delta_3<0$ which
	     parallels the case   $\delta_2>0$ for Fig. 2. This is
	     again as expected. An analysis similar to the above 
	     for the case when $\delta_1=\delta_2=\delta_3=0$ and 
	     $\delta_4$ takes on a range of values is given in Fig. 4. 
	     The result of non-universalities is qualitatively similar
	     to that of  Fig. 3 as expected. 
	     We note, however, that the parameter space allowed by
	     the non-universalities  in the third  generation would  be
	     somewhat different and so one does not have a complete
	     correspondence between the non-universalities in the 
	     Higgs sector~~(i.e., $\delta_2$)  and those in the third 
	     generation sector (i.e., $\delta_3$ and $\delta_4$).\\
\noindent
{\bf 7. Effects of Grand Unification On Non-Universal Parameters }\\
In the previous section, the parameters describing non-universal 
soft SUSY breaking masses, i.e., $\delta_1, .. ,\delta_4$, were viewed
as independent quantities. This would be the case  if the scale of SUSY
breaking occured {\it below} $M_G$, since there only the SM gauge group
is assumed to hold. However, if the SUSY breaking scale occurs above
$M_G$(e.g., at the string scale $M_{str}$), then the soft breaking masses
at $M_G$ must obey constraints imposed by the grand unification group G.
We consider here this situation for the case where G contains SU(5) as a
subgroup. This includes G=SU(5), SO(N) for N$\geq 10$ or E(6) which 
essentially include all the grand unification groups that have been 
examined in significant detail. 

For this class of groups, we may characterize the light matter at $M_G$ in
terms of the SU(5) qunatum numbers, i.e., matter occurs in $10_a$
and $\bar 5_a$ representations (a=1,2,3 is a generation index) and 
the light Higgs doublet in a 5 and $\bar 5$ representation. One has then 
for the sfermions\cite{flipped} 
\begin{eqnarray}
10_a=(q_a\equiv (\tilde u_{La}, \tilde d_{La}); u_a\equiv \tilde u_{Ra};
e_a\equiv\tilde e_{Ra}) \\
\bar 5_a=(l_a\equiv(\tilde \nu_{La},\tilde e_{La}); d_a\equiv\tilde d_{Ra})
\end{eqnarray}
The soft breaking masses have the form 

\begin{eqnarray}
 m_{10a}^2=m_0^2(1+\delta_a^{10});~~m_{\bar 5a}^2=m_0^2(1+\delta_a^{\bar 5})
\end{eqnarray}
with the Higgs masses obeying Eq. (7). There are therefore 8 independent 
soft breaking parameters for this general case. ( Eqs. (7, 43) contains 9
parameters $\delta_a^{10}$, $\delta_a^{\bar 5}$, $\delta_{1,2}$, $m_0$,
but actually one is redundant.). Thus we have the relations 
\begin{equation}
\delta_{q_a}=\delta_{u_a}=\delta_{e_a}=\delta_a^{10};~~\delta_{l_a}=\delta_{d_a}
=\delta_a^{\bar 5}
\end{equation}
For higher gauge groups, there can be additional relations. (For example, for
SO(10) with non-universalities in the third generation, some models of 
symmetry breaking give\cite{kawamura} $\delta^{\bar 5}$=$\delta_2$-$\delta_1$.)

One may impose further group constraints on the above parameters. Thus the 
suppression of FCNC suggests a near degeneracy in the first two generaitons,
which occurs naturally in a SU(2)$_H$ model where these generations 
are put into a doublet (d) representation of SU(2)$_H$ and the third 
generation  into  a singlet (s) representation\cite{dine}. Then Eqs. (45)
reduce to 
\begin{eqnarray}
m_{10}^2=m_0^2(1+\delta_{(s)}^{10});~~m_{10_a}^2=m_0^2(1+\delta_{(d)}^{10}), 
a=1,2\\
m_{\bar 5}^2=m_0^2(1+\delta_{(s)}^{\bar 5}); 
~~m_{\bar 5_a}^2= m_0^2(1+\delta_{(d)}^{\bar 5}), a=1,2
\end{eqnarray}
where quantities without generation indices are SU(2)$_H$ singlets. This 
model would have 6 independent parameters. 

While the FCNC constraints do not imply that the first two generation
soft mass of the 10 representation is degenerate  with that of the $\bar 5$
representation, a further simplification occurs if one assumes that  this
is the case, i.e., one lets \
\begin {equation}
\delta_{(d)}^{\bar 5}=\delta_{(d)}^{10}\equiv \delta_{(d)}
\end{equation}
One can then absorb the $(1+\delta_{(d)})$ factor into the $m_0^2$ as the
common mass for a=1,2, and use that as the reference mass for the third
generation and Higgs masses, reducing the number of parameters to 5:
$\delta_{(s)}^{10}$, $\delta_{(s)}^{\bar 5}$, $\delta_{1,2}$ and
$m_0$. In the notation of Sec. 7, this is equivalent to 
non-universalities specified by 
\begin{equation}
\delta_3=\delta_4;~\delta_1;~\delta_2;~\delta_5
\end{equation}
where $\delta_5\equiv \delta_{(s)}^{\bar 5}=\delta_{\tilde b_R}=\delta_{l_3}$.
The parameter $\delta_5$ does not  enter directly into the event rate R
calculation or into the Landau pole or $b\rightarrow s+\gamma$ analyses.
It does, however, affect R indirectly in that it will enter into the 
relic density calculation and affect the determination of the 
parameter space that obeys Eq. (1). However, there it enters only into
the t-channel poles which generally make small contributions to the 
$\chi_1^0$ annihilation cross section. We discuss now the  effect of the
 model of Eq. (50) on event rates. For illustration we consider two cases 
 with $\delta_i$(i=3,4) chosen at the extreme values under the reasonable
constraint $\delta_i\leq 1$. These are
 case(a): $\delta_3$=$\delta_4$=1,  and case(b):
 $\delta_3$=$\delta_4$=-1(and $\delta_5$ set to zero). For case(a) we find
 from Eqs. (13), (20)  and (22) using $D_0\simeq 0.27$ that 
 	     \begin{equation}
	      \Delta_{H_2}\simeq -0.73+0.64\delta_2
	      \end{equation}
	     \begin{equation}
	       \Delta_{\tilde Q}\simeq 0.76 -0.12\delta_2
	      \end{equation}
	     \begin{equation}
	      \Delta_{\tilde U}\simeq 0.51 -0.24\delta_2
	      \end{equation}
For $\delta_2$ in the range (1,-1) one finds that $\Delta_{H_2}$ lies in 
the range (-0.1,-1.37) while $\Delta_{\tilde Q}$ lies in the range
(0.64, 0.88) and $\Delta_{\tilde U}$ lies in the range (0.27,0.76).
Because of the non-universalities arising from $\delta_3=\delta_4 > 0$
this case is again different from the cases we have discussed before.
However, we can still draw some rough comparisons with the previous cases. 
Thus, for example, we note that since the effective 
value of $\delta_2$ in this case is negative the minimum of the event rates 
should be similar to the minimum dotted curve of Fig. 2 where $\delta_2$
was also negative. The result of 
the analysis is presented  in Fig. 5. We see that indeed the minimum of the 
event rates here is similar to the minimum of the event rates in Fig. 2 as
expected. In general, the effect of non-universalities is not very large 
here.
 A similar analysis holds for case(b). Here again
 using  Eqs. (13), (20), (22) and  $D_0\simeq 0.27$ we find that 
 	     \begin{equation}
	      \Delta_{H_2}\simeq 0.73+0.64\delta_2
	      \end{equation}
	     \begin{equation}
	       \Delta_{\tilde Q}\simeq -0.76 -0.12\delta_2
	      \end{equation}
	     \begin{equation}
	      \Delta_{\tilde U}\simeq -0.51 -0.24\delta_2
	      \end{equation}
For $\delta_2$ in the range (1,-1) one finds $\Delta_{H_2}$ in 
the range (0.1,1.37), $\Delta_{\tilde Q}$ in the range
(-0.64, -0.88), and $\Delta_{\tilde U}$ in the range (-0.27,-0.76).
We note that  these ranges  are exact mirrors of the ranges for case(a) 
but with a negative  sign. Of course, the full dynamics of this system is very
complicated since the allowed parameter space is affected differently 
from previous cases because of the specific nature of non-universalities 
here. However, as in case(a) we can glean some understanding by a comparison
with the previous analyses. We note that  the effective value of  $\delta_2$
is positive here. Thus one  expects that the minimum event rates
might show the same  pattern as the positive $\delta_2$ case of Fig. 2. 
The result of the analysis is exhibited in Fig. 6. We find that indeed
 the minima curves of Fig. 6 have a behavior similar to the dashed  curve of
Fig. 2  as expected. The remarkable feature here is the increase by a 
factor of 10 in the maximum
event rates when $\delta_2$ is positive. This is due to the fact that 
$\delta_2$,$\delta_3$, $\delta_4$ now act together to reduce $\mu^2$ 
(see Eqs. (23) and (24)) which increases the maximum event rates$^{38}$.

\noindent
{\bf 8. Effects of More Accurate $\Omega h^2$ Determination}\\
As mentioned in the Introduction the next round of satellite 
experiments are expected to determine the Hubble parameter
to within O(10\%) or better accuracy. 
Such an accurate  determination of the 
 Hubble parameter will put more stringent bounds on the neutralino 
  relic density  $\Omega_{\tilde\chi_1}h^2$
 and hence more stringently constrain the SUGRA 
parameter space. We investigate here the effects of these more
stringent constraints on the neutralino dark matter event rates.
 For specificity we  consider a narrow band of $\Omega_{\tilde\chi_1} h^2$ 
 in the range
\begin{equation}
  0.225<\Omega_{\tilde\chi_1} h^2<0.275
\end{equation}
 Fig. 7 gives the maximum and the minimum event rates as a function of the
 $\tilde\chi_1$ mass. A comparison of Figs. 2 and 7 shows that a 
     narrower range $\Omega_{\tilde\chi_1} h^2$ 
      more sharply  constrains the SUGRA parameter space
   	and the allowed band of event rates shrinks.
   	The allowed range of the neutralino mass is also
   	more sharply constrained since the $m_{\tilde\chi_1}$ mass is 
   	constrained to lie below $\approx 90 $ GeV, as above this mass
   	$\Omega h^2$ exceeds the upper limit of Eq. (57).  Since the 
   	neutralino mass and the gluino mass scale one also deduces an
   	upper limit on the gluino mass of about 650 GeV. Thus a 
   	precise measurement of  $\Omega_{\tilde\chi_1} h^2$
   	 will put an interesting upper bound on the gluino mass which
   	 would be testable at accelerators such as the upgraded Tevatron 
   	 and the LHC. In general the overall effect of the more stringent 
 constraints on  $\Omega_{\tilde\chi_1} h^2$ 
 is to eliminate a part of the parameter space allowed by Eq. (1) and 
 thus limit more severely the permissible domain of the event rates.\\
\noindent
{\bf 9. Conclusions}\\
In this paper we  have investigated the effects of  non-universalities 
in the Higgs secotor and in the third generation squark 
sector on the low energy parameters. We have shown that the Higgs 
sector non-universalities and the non-universalities in the third 
generation  squark sector are highly coupled and we 
have exhibited analytic  forms of  their effects on the 
parameters $\mu$, $m_A$ and the third generation squark masses
at low energy. 
Next we investigated the effects of these non-universalities
on the event rates in the scattering of neutralinos by nuclei. 
Here  we found that the minimum event rates in the neutralino mass region 
$30 ~GeV\leq m_{\tilde\chi_1}\leq ~65 GeV$ can be increased or decreased by a  
factor
of O(10) depending on the sign of the non-universality. In contrast,
in the neutralino mass region $m_{\tilde\chi_1}\geq 65 ~GeV$ the minimum 
and the maximum event rates are not appeciably affected. 
 These results  have important implications for dark matter detectors.
 Supergravity unified models with universal boundary conditions for soft
 SUSY breaking parameters give event rates which lie in a wide range
  R= O$(1-10^{-5})$ events/kg da Thus while the current generation of
 dark matter  detectors can sample a part of the parameter space,
 one needs far more sensitive detectors (more sensitive by a factor of
 10$^{3}$) to  sample the entire parameter space of the model.
However, inclusion of non-universalities shows that for certain
signs of the non-universality, the minimum event rates can increase
by a factor of 10-100 in the region $m_{\tilde\chi_1}\leq 65 GeV$. 
These models can be completely tested in this mass region if the
detector sensitivity could be enhanced by a factor of about 10$^2$.
Such an enhancement does not seem out of reach with new techniques 
currently being discussed\cite{cline}. Another interesting 
result seen was the appearance of large event rates, up to 10 event/kg da
for the case when $\delta_3=\delta_4=-1$. This case looks
very interesting from the view point of experimental detection of dark matter. 
We have also analysed  the
implications of more stringent constraints on $\Omega h^2$
which is expected from the next round of satellite experiments, 
 and found that while the event rates typically become
smaller a significant part of the parameter space yields event rates
which are  within reach of dark matter experiments currently
being planned.


\noindent
{\bf Acknowledgments}\\
This research was supported in part by NSF grant numbers 
PHY-96020274 and PHY-9411543.

\noindent
{\bf Appendix A}\\
All the scalar sparticle masses receive contributions from the Tr(Y$m^2$) 
term. Corrections to the first two generations of sparticle masses 
are given by 

\begin{eqnarray}
\Delta m_{\tilde u_{iL}}^2=-\frac{1}{5}S_0p=\Delta m_{\tilde d_{iL}}^2,
~~~~\Delta m_{\tilde e_{iL}}^2=\frac{3}{5}S_0p=\Delta m_{\tilde \nu_{i}}^2\\
\Delta m_{\tilde u_{iR}}^2=\frac{4}{5}S_0p=-2\Delta m_{\tilde d_{iR}}^2,
\Delta m_{\tilde e_{iR}}^2=-\frac{6}{5}S_0p
\end{eqnarray}
where $S_0$ and p are defined in the text.
Under the assumption that the squarks and slepton  masses are universal 
at the GUT scale for the first two generations, but with  
non-universaliites in the Higgs sector and in the third generation
we can deduce the following sum rule for the Tr(Y$m^2$) term:
\begin{equation}
p Tr(Ym^2)_0=-\frac{5}{6}sin^2\theta_W cos2\beta M_Z^2+\frac{1}{4}
( m_{\tilde e_{iL}}^2 +2 m_{\tilde u_{iR}}^2)-\frac{1}{4}
(m_{\tilde e_{iR}}^2 +m_{\tilde d_{iR}}^2 + m_{\tilde d_{iL}}^2)
\end{equation}
Here $Tr(Ym^2)_0$ is the value of $Tr(Ym^2)$ at the GUT scale while the
right hand side is evaluated at the electro-weak scale.
We see from above that if the masses of the first two generations of 
squarks and sleptons are known with enough precision, one can determine 
the value of the trace anomly term at the GUT scale. For the case 
when $Tr(Ym^2)_0$ vanishes the above sum rule reduces to the one 
given in Ref. (56).

$m_{H_2}^2$,  $m_{\tilde U}^2$ and  $m_{\tilde Q}^2$ satisfy 
 the following set of coupled equations 
\begin{equation}
\frac{dm_{H_2}^2}{dt}=-3Y_t\Sigma_t+(3\tilde\alpha_2M_2^2+\frac{3}{5}
\tilde\alpha_1M_1^2)+\gamma_1\tilde\alpha_1S
\end{equation}
\begin{equation}
\frac{dm_{\tilde U}^2}{dt}=-2Y_t\Sigma_t+(\frac{16}{3}\tilde\alpha_3M_3^2
+\frac{16}{15}\tilde\alpha_1M_1^2)+\gamma_2\tilde\alpha_1S
\end{equation}
\begin{equation}
\frac{dm_{\tilde Q}^2}{dt}=-Y_t\Sigma_t+(\frac{16}{3}\tilde\alpha_3M_3^2
+ 3\tilde\alpha_2M_2^2+\frac{1}{15}\tilde\alpha_1M_1^2)
+\gamma_3\tilde\alpha_1S
\end{equation}
where $\Sigma_t=(m_{H_2}^2+m_{\tilde Q}^2+m_{\tilde U}^2+A_t^2)$, 
and $\gamma_1$=-$\frac{3}{10}$, $\gamma_2$=$\frac{2}{5}$
and $\gamma_3$=-$\frac{1}{10}$ and S is the running parameter with
S(0)=$S_0$. We show now
that the evolution of the trace anomaly term is not affected by the top
Yukawa coupling. To see this it is  best to go to a decoupled set
of equations using the functions $e_i(t)$ given by 
\begin{eqnarray}
m_{H_2}^2=3e_1+e_2+e_3\\
m_{\tilde U}^2=2e_1-2e_2\\
m_{\tilde Q}^2=e_1+e_2-e_3
\end{eqnarray}
It is easily seen that the evolution of $e_2$, and $e_3$ does 
not involve the Yukawa coupling and only the evolution of $e_1$ 
 does. One has 
\begin{equation}
\frac{de_1}{dt}=-6Y_te_1-Y_tA_t^2+(\frac{16}{9}\alpha_3M_3^2
+\alpha_2M_2^2+\frac{13}{45}\alpha_1M_1^2)+\frac{1}{6}
(\gamma_1+\gamma_2+\gamma_3)\tilde\alpha_1S
\end{equation}
However, since 
\begin{equation}
(\gamma_1+\gamma_2+\gamma_3)=0
\end{equation}
 one finds that the 
evolution of the anomaly terms in Eqs. (61)-(63)  are not affected by the top 
Yukawa coupling and are thus not sensitive to the Landau pole effects. 
The bottom squark mass matrix is given by 
\begin{equation}
\left(
{{ {m_{\tilde b_L}^2}\atop{-m_{b} (A_b + \mu tan \beta)}}
{{-m_b (A_b + \mu tan\beta)} \atop {m_{\tilde b_R}^2}   }}
\right)
\end{equation}
where
\begin{eqnarray}
m_{\tilde b_L}^2=m_{\tilde Q}^2+ m_b^2+ 
(-\frac{1}{2}+\frac{1}{3}sin^2\theta_W)M_Z^2cos2\beta
\end{eqnarray}
and where $m_{\tilde Q}^2$ is defined in the text. \\

\noindent
{\bf Figure Captions}\\

\noindent
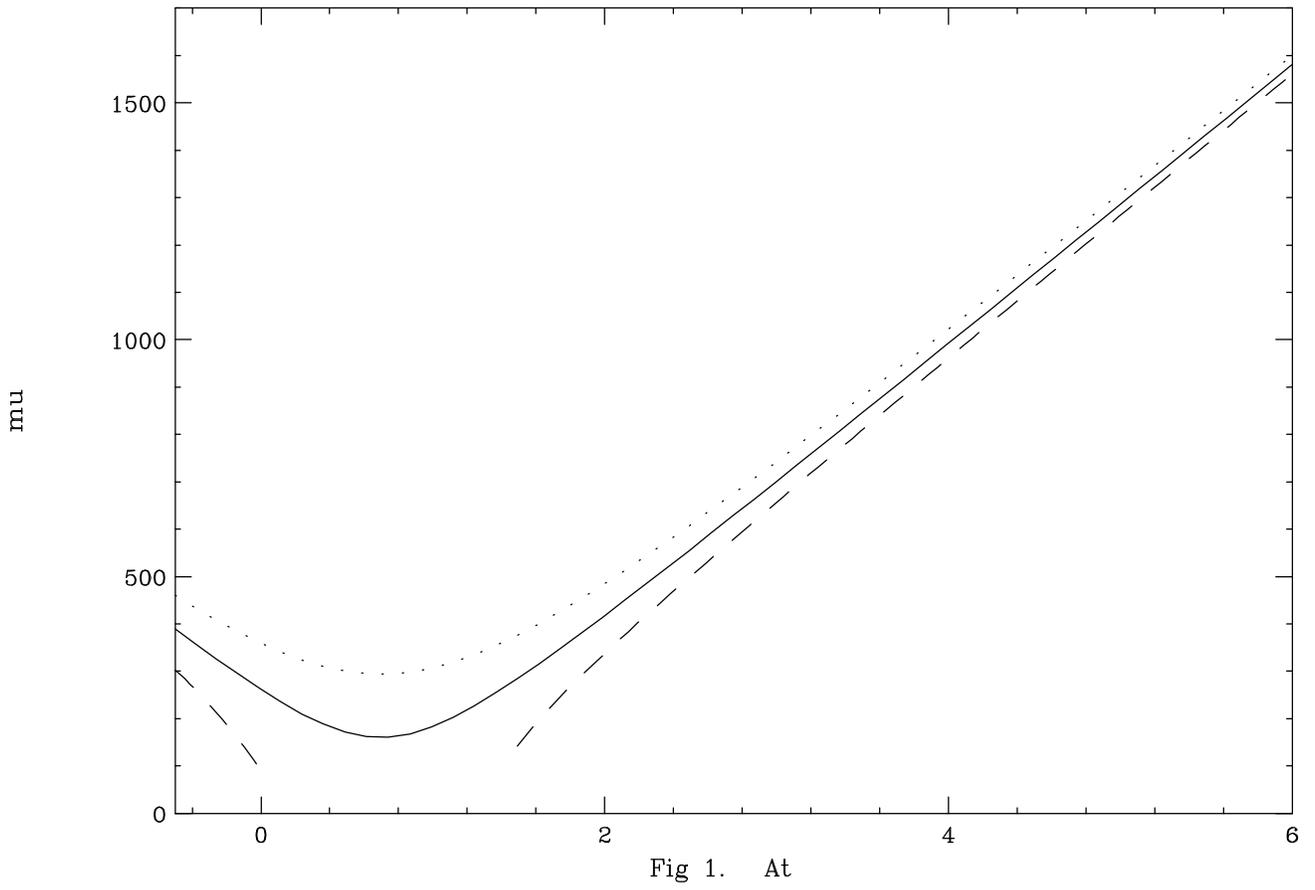
Fig.1: $\mu$ vs $A_t$  for the case when $\delta_3=0=\delta_4$,
$m_0=300$ GeV, $m_{\tilde g}=350$ GeV,  and
(i)$\delta_1=0=\delta_2$(solid), (ii) $\delta_1=-1=-\delta_2$(dashed),
and (iii)$\delta_1=1=-\delta_2$(dotted).   
   Here  $A_R$ vanishes for $A_t/m_0\approx 0.7$.\\

\noindent
Fig. 2: Maximum and minimum of event rates/kg da for xenon  for $\mu>0$
for the case
 when $\delta_3=0=\delta_4$ and 
 (a)$\delta_1=0=\delta_2$(solid), (b)$\delta_1=1=-\delta_2$(dotted), 
 and (c)$\delta_1=-1=-\delta_2$ (dashed)
  when $0.1<\Omega_{\tilde\chi_1} h^2<0.4$, and $m_t$=175 GeV. \\

\noindent
Fig. 3: Maximum and minimum of event rates/kg da for xenon for the case
when  $\delta_1=\delta_2=\delta_4$=0 and 
  (a)$\delta_3=0$(solid), (b)$\delta_3$=1(dotted), and (c)$\delta_3$=-1
   (dashed) when $0.1<\Omega_{\tilde\chi_1} h^2<0.4$, and $m_t$=175 GeV. \\

\noindent
Fig. 4: Maximum and minimum of event rates/kg da for xenon for the case
 when $\delta_1=\delta_2=\delta_3$=0 and
  (a)$\delta_4=0$(solid), (b)$\delta_4$=1(dotted), and (c)$\delta_4$=-1
   (dashed) when $0.1<\Omega_{\tilde\chi_1} h^2<0.4$, and $m_t$=175 GeV. \\

\noindent
Fig. 5: Maximum and minimum event rates when $\delta_3=\delta_4=1$, 
$\delta_5=0$ and (a) $\delta_1=0=\delta_2$(solid), (b) $\delta_1$=1=
-$\delta_2$ (dotted), and  $\delta_1$=-1=-$\delta_2$ (dashed) 
when $0.1<\Omega_{\tilde\chi_1} h^2<0.4$, and $m_t$=175 GeV. \\

\noindent
Fig. 6: Same as Fig. 5 when  $\delta_3=\delta_4=-1$.\\

\noindent
Fig. 7: Maximum and minimum of event rates/kg da for xenon for the case 
when $\delta_3=\delta_4$=0 and 
(i) $\delta_1=0=\delta_2$(solid),
(ii)$\delta_1=-1=-\delta_2$(dashed), and 
(iii)$\delta_1=1=-\delta_2$(dotted)
 when $0.225<\Omega_{\tilde\chi_1} h^2<0.275$, and $m_t$=175 GeV. 
\newpage

\begin{center} \begin{tabular}{|c|c|c|c|c|}
\multicolumn{5}{c}{}\\
\hline
\hline
$m_{\tilde\chi_1}(GeV)$   & 41.5 & 41.5  &52.4&52.4   \\   
\hline
  & ~~~~~$R_{Min}$~~~~~ & ~~~~~$R_{Max}$~~~~~ & ~~~~~$R_{Min}$
 ~~~~~ & $R_{Max}$\\
\hline
He &  6.97E-4 & 3.43E-3 & 6.89E-4 & 2.49E-3 \\
CaF2 &  8.33E-4 & 4.26E-3      
 &  9.02E-4 & 2.25E-2 \\
GaAs &  1.46E-4 & 1.40E-3  & 1.58E-4 & 8.14E-2 \\
 Ge & 1.08E-4 & 1.23E-3  &  1.11E-4 & 8.33E-2  \\
 NaI & 1.18E-4 & 1.68E-3  & 1.14E-4 & 0.12    \\
 Pb & 1.82E-4 & 2.83E-3  & 1.68E-4 & 0.21  \\
\hline
\hline 
 $m_{\tilde\chi_1}$(GeV)& 61.4& 61.4 & 71.2 & 71.2 \\   
\hline
  & ~~~~~$R_{Min}$~~~~~ & ~~~~~$R_{Max}$~~~~~ & ~~~~~$R_{Min}$
 ~~~~~ & $R_{Max}$\\
\hline
 He & 2.95E-5 & 9.35E-3 &  3.05E-7 & 4.28E-3 \\
CaF2 &  3.53E-4 & 5.48E-2 &   1.51E-4 & 2.45E-2\\
GaAs &  1.32E-3 & 0.17  &   6.41E-4 & 7.89E-2  \\
Ge &  1.35E-3 & 0.18 &  6.57E-4 & 8.04E-2 \\    
NaI &  2.02E-3 & 0.27  &  9.65E-4 & 0.11     \\  
 Pb &  3.38E-3 & 0.45  &  1.58E-3 & 0.19     \\    
\hline
\hline
$m_{\tilde\chi_1}$(GeV) & 80.7 & 80.7 &  117.5 &  117.5   \\    
\hline
  & ~~~~~$R_{Min}$~~~~~ & ~~~~~$R_{Max}$~~~~~ & ~~~~~$R_{Min}$
 ~~~~~ & $R_{Max}$\\
\hline
 He & 5.23E-6 & 2.31E-3 & 1.01E-4 & 1.87E-4    \\
 CaF2 & 8.95E-5&  1.27E-2 &  2.29E-4 & 8.19E-4 \\
GaAs &  3.47E-4 & 3.99E-2 &3.01E-4  &2.22E-3 \\
Ge &  3.55E-4 & 4.06E-2 & 2.96E-4 & 2.25E-3 \\
NaI &  5.14E-4 & 5.86E-2 & 4.03E-4 & 3.11E-3 \\
Pb &   8.32E-4 & 9.45E-2  & 6.27E-4 & 4.86E-3   \\
\hline 
\end{tabular} 
\end{center}

\noindent
Table 1: Minimum and maximum of event rates/kg da are given for the case 
$\delta_1=0=\delta_2$ and $\mu>0$ for several neutralino masses for the
 target materials He, CaF2, GaAs, Ge, NaI, and Pb. 
\newpage

\begin{center} \begin{tabular}{|c|c|c|c|c|}
\multicolumn{5}{c}{}\\
 \hline
\hline
 $m_{\tilde\chi_1}$ (GeV)  & ~~~~41.6~~~~ & ~~~~41.6~~~~ &  ~~~~53.9~~~~&  53.9 \\
\hline
  & ~~~~~$R_{Min}$~~~~~ & ~~~~~$R_{Max}$~~~~~ & ~~~~~$R_{Min}$
 ~~~~~ & $R_{Max}$\\
\hline
He   &3.57E-5 & 3.61E-5 & 7.88E-6 & 1.91E-3 \\
CaF2 &  4.48E-5 &  1.034E-3 & 1.07E-5 & 1.79E-2 \\
GaAs & 1.62E-5 & 4.10E-3 &  3.05E-6 &  6.53E-2\\
Ge & 1.45E-5 & 4.21E-3 & 2.53E-6 & 6.68E-2\\
NaI & 1.99E-5 & 6.56E-3 & 3.20E-6 & 0.10  \\
Pb & 3.38E-5 & 1.15E-2 & 5.13E-6 & 0.17   \\
   \hline
\hline 
$m_{\tilde\chi_1}(GeV)$   & 61.0 & 61.0  & 70.1 & 70.1   \\  
\hline
  & ~~~~~$R_{Min}$~~~~~ & ~~~~~$R_{Max}$~~~~~ & ~~~~~$R_{Min}$
 ~~~~~ & $R_{Max}$\\
\hline
 He & 3.18E-4 & 1.09E-2 &    1.48E-4 & 5.05E-3 \\
 CaF2 & 6.96E-4 &  6.06E-2   &   3.52E-4 & 2.77E-2\\
GaAs & 1.16E-3 & 0.19 & 6.30E-4 & 8.81E-2\\
Ge & 1.17E-3 & 0.19  & 6.32E-4 & 8.98E-2\\
NaI & 1.72E-3 & 0.29 & 9.18E-4 & 0.13    \\
Pb & 2.86E-3 & 0.49  & 1.50E-3 & 0.21    \\
\hline
\hline   
  $m_{\tilde\chi_1}(GeV)$ &  80.9&  80.9 & 117.0 & 117.0 \\      
\hline
  & ~~~~~$R_{Min}$~~~~~ & ~~~~~$R_{Max}$~~~~~ & ~~~~~$R_{Min}$
 ~~~~~ & $R_{Max}$\\
\hline
He & 1.02E-5 & 2.15E-3  &    1.10E-4 & 1.80E-4 \\
 CaF2 & 8.87E-5 & 1.14E-2 & 2.28E-4 & 7.85E-4\\
GaAs & 3.14E-4 & 3.53E-2 & 2.36E-4 & 2.12E-3\\
Ge & 3.20E-4 & 3.60E-2 & 2.29E-4 & 2.14E-3\\
NaI & 4.63E-4 & 5.19E-2 & 3.09E-4 & 2.97E-3\\
Pb & 7.49E-4 & 8.37E-2 & 4.78E-4 & 4.63E-3\\
\hline 
\end{tabular} 
\end{center}

\noindent
Table 2: Same as Table 1 except that $\delta_1=1=-\delta_2$   

\newpage

\begin{center} \begin{tabular}{|c|c|c|c|c|}
\multicolumn{5}{c}{}\\
 \hline
$m_{\tilde\chi_1}$(GeV) &~~~~50.3~~~~ &~~~~50.3~~~~ & ~~~~54.6~~~~&  54.6   \\   
\hline
  & ~~~~~$R_{Min}$~~~~~ & ~~~~~$R_{Max}$~~~~~ & ~~~~~$R_{Min}$
 ~~~~~ & $R_{Max}$\\
\hline
He &  6.61E-3 & 1.50E-2 & 1.14E-4 & 4.36E-2 \\
 CaF2 & 8.90E-3 &  0.10 &7.33E-4 & 0.28   \\
GaAs & 3.11E-3 & 0.36 & 2.47E-3 & 0.95    \\
Ge & 2.72E-3 &  0.37 & 2.52E-3 & 0.97    \\
NaI & 3.64E-3 & 0.57 &3.82E-3 &  1.47    \\
Pb & 5.99E-3 & 0.97 & 6.47E-3 &  2.49    \\
  \hline
  \hline
$m_{\tilde\chi_1}$(GeV)   & 61.2  & 61.2 &   70.9 &   70.9    \\ 
\hline
  & ~~~~~$R_{Min}$~~~~~ & ~~~~~$R_{Max}$~~~~~ & ~~~~~$R_{Min}$
 ~~~~~ & $R_{Max}$\\
\hline
He & 3.71E-5 & 2.80E-2 & 3.74E-6 & 7.30E-2\\
CaF2 & 4.09E-4 & 0.14 & 1.81E-4 & 0.16    \\
GaAs & 1.52E-3 & 0.45 & 7.46E-4 & 0.29    \\
Ge & 1.55E-3 & 0.46 & 7.65E-4 & 0.29   \\
NaI & 2.32E-3 & 0.68 & 1.12E-3 & 0.43    \\
Pb & 3.88E-3  & 1.14 & 1.84E-3 & 0.71    \\
 \hline  
\hline
$m_{\tilde\chi_1}(GeV)$   &  80.5  &  80.5    &117.5  &117.5  \\  
\hline
  & ~~~~~$R_{Min}$~~~~~ & ~~~~~$R_{Max}$~~~~~ & ~~~~~$R_{Min}$
 ~~~~~ & $R_{Max}$\\
\hline
He & 1.65E-6 & 8.20E-2 & 1.04E-4 & 8.02E-2\\ 
CaF2 & 9.40E-5 &  0.18 & 2.37E-4 & 1.10E-2    \\
GaAs & 3.88E-4 & 0.33 & 3.12E-4 & 1.89E-2\\
Ge & 3.98E-4 & 0.33 & 3.07E-4 & 1.23E-2\\
NaI & 5.76E-4 & 0.49 & 4.18E-4 & 1.14E-2\\
Pb & 9.33E-4 & 0.81 & 6.50E-4 & 1.57E-2\\  
\hline 
\end{tabular} 
\end{center}

\noindent
Table 3: Same as Table 1 except that $\delta_1=-1=-\delta_2$   

\newpage
 \begin{center} \begin{tabular}{|c|c|c|c|}
\multicolumn{4}{c}{}\\
 \hline
 $A_t$ & $\delta_1=0=\delta_2$ & $\delta_1=1=-\delta_2$&$\delta_1=-1=-\delta_2$\\   
\hline
  & $R_{Min} ~~R_{Max}$ & $R_{Min}~~R_{Max}$ & $R_{Min}~~R_{Max}$ \\
-0.5 & - & 3.88E-6~~2.83E-3 & - \\
 0 &  5.97E-5~~1.28E-2&6.43E-6~~1.53E-2 &   4.40E-3~~0.19   \\
 1.0 &     2.28E-3~~0.26 & 1.10E-5~~7.14E-2 & 1.15E-2~~1.82   \\    
  2.0 & 2.73E-3~~0.36 &1.80E-3  ~~0.20  & 5.14E-3   ~~1.22  \\    
  3.0 &       6.83E-4   ~~0.52 &  5.82E-4   ~~0.36    &  8.14E-4   ~~0.79   \\    
  4.0 &      3.19E-4 ~~0.53 & 2.90E-4 ~~0.43  & 3.52E-4~~0.84   \\    
  5.0 &      3.10E-4 ~~0.56 &  3.53E-4 ~~0.45    &3.42E-4~~0.71   \\    
  6.0 &   2.14E-4~~0.46 & 3.84E-4 ~~0.37   & 2.22E-4~~0.58  \\    
 7.0  &    1.82E-4 ~~0.21 & 1.66E-4 ~~0.17    &  2.01E-4 ~~0.25 \\    
\hline 
\end{tabular} 
\end{center}

\noindent
Table 4: A comparison of the minimum and maximum  event rates/kg da 
 for xenon for the cases $\delta_1=0=\delta_2$,
~$\delta_1=1=-\delta_2$, 
 and $\delta_1=-1=-\delta_2$, and $\mu>0$ as a function of $A_t$.\\

\begin{center} \begin{tabular}{|c|c|c|c|}
\multicolumn{4}{c}{}\\
 \hline
 $A_t$ & $\delta_1=0=\delta_2$ & $\delta_1=1=-\delta_2$&$\delta_1=-1=-\delta_2$\\   
\hline
   & $R_{Min}~~R_{Max}$  & $R_{Min}~~R_{Max}$  & $R_{Min}~~R_{Max}$  \\
    -0.5& 4.65E-6 ~~6.40E-3 & 2.55E-6~~1.33E-2 &4.65E-6~~1.24E-4 \\
   0.0 &  4.65E-6~~8.15E-3 & 2.22E-6 ~~2.33E-2 &  4.65E-6~~1.73E-2\\
   1.0 & 4.66E-6 ~~1.48E-3  &4.67E-6~~2.65E-3 & 4.67E-6~~9.04E-4\\
   2.0  &  4.67E-6~~6.18E-4 &4.67E-6 ~~3.65E-3 & 4.67E-6~~ 6.21E-4\\
  3.0 &  4.67E-6~~5.55E-4 & 4.67E-6~~5.49E-4 &4.67E-6 ~~5.60E-4 \\
  4.0 &  4.67E-6 ~~4.40E-4 & 4.67E-6~~ 4.36E-4 &4.67E-6~~4.44E-4 \\
   5.0 &    4.67E-6~~3.49E-4 &4.67E-6~~ 3.46E-4 & 4.67E-6~~1.01E-3 \\
    6.0   &  4.67E-6~~3.09E-4 &4.67E-6~~2.77E-4 & 4.67E-6 ~~3.11E-4\\
    7.0  & 4.67E-6~~3.17E-4  &4.67E-6~~ 3.15E-4 &  4.67E-6~~1.0E-2  \\      
\hline 
\end{tabular} 
\end{center}

\noindent
Table 5: Same as Table 4 except that $\mu<0$. 

\newpage

\end{document}